\documentclass[prl,twocolumn,superscriptaddress,floatfix,noshowpacs,10pt]{revtex4-1}%
\usepackage{graphicx,bm,times}
\usepackage{amsmath}
\usepackage{amsfonts}
\usepackage{amssymb}

\begin{document}

\title{Dichotomy between the hole and electrons behavior
in the multiband FeSe \\ probed by ultra high magnetic fields }

\affiliation{Department of Physics, Kyoto University, Sakyo-ku, Kyoto 606-8501, Japan}

\author{M.\,D. Watson}
\affiliation{Clarendon Laboratory, Department of Physics,
University of Oxford, Parks Road, Oxford OX1 3PU, UK}

\author{T. Yamashita}
\affiliation{Department of Physics, Kyoto University, Sakyo-ku, Kyoto 606-8501, Japan}

\author{S. Kasahara}
\affiliation{Department of Physics, Kyoto University, Sakyo-ku, Kyoto 606-8501, Japan}

\author{W. Knafo}
\affiliation{Laboratoire National des Champs Magn\'etiques Intenses, 31400 Toulouse, France}

\author{M. Nardone}
\affiliation{Laboratoire National des Champs Magn\'etiques Intenses, 31400 Toulouse, France}

\author{J. B\'{e}ard}
\affiliation{Laboratoire National des Champs Magn\'etiques Intenses, 31400 Toulouse, France}

\author{F. Hardy}
\affiliation{Institut fur Festk\"{o}rperphysik, Karlsruhe Institute of Technology, 76021 Karlsruhe, Germany}

\author{A.~McCollam}
\affiliation{High Field Magnet Laboratory, Institute
for Molecules and Materials, Radboud University, 6525 ED Nijmegen, The Netherlands}

\author{A. Narayanan}
\affiliation{Clarendon Laboratory, Department of Physics,
University of Oxford, Parks Road, Oxford OX1 3PU, UK}

\author{S.\,F. Blake}
\affiliation{Clarendon Laboratory, Department of Physics,
University of Oxford, Parks Road, Oxford OX1 3PU, UK}

\author{T. Wolf}
\affiliation{Institut fur Festk\"{o}rperphysik, Karlsruhe Institute of Technology, 76021 Karlsruhe, Germany}

\author{A.\,A. Haghighirad}
\affiliation{Clarendon Laboratory, Department of Physics,
University of Oxford, Parks Road, Oxford OX1 3PU, UK}

\author{C. Meingast}
\affiliation{Institut fur Festk\"{o}rperphysik, Karlsruhe Institute of Technology, 76021 Karlsruhe, Germany}

\author{A.\,J. Schofield}
\affiliation{School of Physics and Astronomy, University of Birmingham,
Edgbaston, Birmingham B15 2TT, UK}

\author{H. von L\"ohneysen}
\affiliation{Institut fur Festk\"{o}rperphysik, Karlsruhe Institute of Technology, 76021 Karlsruhe, Germany}

\author{Y. Matsuda}
\affiliation{Department of Physics, Kyoto University, Sakyo-ku, Kyoto 606-8501, Japan}

\author{A.\,I. Coldea}
\email[corresponding author: ]{amalia.coldea@physics.ox.ac.uk}
\affiliation{Clarendon Laboratory, Department of Physics,
University of Oxford, Parks Road, Oxford OX1 3PU, UK}

\author{T. Shibauchi}
\email[corresponding author: ]{shibauchi@k.u-tokyo.ac.jp}
\affiliation{Department of Advanced Materials Science, University of Tokyo, Kashiwa, Chiba 277-8561, Japan}
\affiliation{Department of Physics, Kyoto University, Sakyo-ku, Kyoto 606-8501, Japan}

\begin{abstract}
Magnetoresistivity $\rho_{xx}$ and Hall resistivity $\rho_{xy}$ in ultra
high magnetic fields up to 88\,T are measured down to 0.15\,K to clarify the multiband electronic structure in high-quality single crystals of superconducting FeSe.
At low temperatures and high fields we observe quantum oscillations
in both resistivity and Hall effect, confirming the multiband Fermi surface with small volumes.
We propose a novel approach to identify the sign of the charge carriers corresponding to a particular cyclotron orbit in a compensated metal from magnetotransport measurements. The observed significant differences in the relative amplitudes of the quantum oscillations between the $\rho_{xx}$ and $\rho_{xy}$ components, together with the positive sign of the high-field  $\rho_{xy}$, reveal that the largest pocket should correspond to the hole band.
The low-field magnetotransport data in the normal state suggest that, in addition to one hole and one almost compensated electron bands, the orthorhombic phase of FeSe exhibits an additional tiny electron pocket with a high mobility.
\end{abstract}
\date{\today}
\maketitle


Uniquely amongst the Fe-based superconductors, the tetragonal $\beta$-FeSe undergoes an orthorhombic distortion at $T_s\approx 90$\,K without an accompanying long-range magnetic ordering at any temperature,
and therefore provides a unique opportunity to probe its unusual electronic behavior from which its superconductivity emerges
below $\sim9$\,K. In particular, recent advances in the high-quality single crystal growth  \cite{Bohmer2013} and several theoretical proposals focused on the frustrated magnetism under nematicity \cite{Wang2015,Glasbrenner2015} have lead to renewed interests of this system as a key material of Fe-based superconductivity.
 Furthermore, it has been suggested that in the superconducting state FeSe may be in the crossover regime between weak-coupling Bardeen-Cooper-Schrieffer and strong-coupling Bose-Einstein condensation limits, which may cause unexpected effects \cite{Kasahara2014}. To understand such intriguing properties of FeSe, the experimental determination of its detailed electronic structure is obviously crucial.

Previous experiments of electronic properties of  FeSe using both ARPES \cite{Nakayama2014,Maletz2013,Shimojima2014,Watson2014} and quantum oscillations \cite{Terashima2014,Audouard2014,Watson2014}, have suggested different interpretations of the electronic structure of FeSe. In this paper by analyzing the magnetotransport data in high magnetic fields,
we reveal the details of the bulk electronic structure and the effects of scattering
from which a more complete picture of the electronic structure of FeSe can be constructed.
We use a novel approach to extract this information by analysing the quantum oscillations measured both in the resistivity $\rho_{xx}$ and Hall effect $\rho_{xy}$ which allows us to deduce the origin of the observed oscillation frequencies in FeSe in fields up to 88\,T. We observe a positive Hall effect suggesting that, in a two band picture, hole carriers have higher  mobility than the (almost compensated) electrons at low temperatures.
 At higher temperatures, the Hall effect changes significantly
 with magnetic field which can only be understood considering a three-band picture,
 with the addition of a small electron band with high mobility.
 Furthermore, the observed large non-saturated magnetoresistance is the consequence of the existence of a small and compensated multiband  electronic structure in FeSe.

{\it Experimental details.}
Samples were grown by the KCl/AlCl$_3$ chemical vapor transport method \cite{Bohmer2013}. 
In-plane transport measurements were performed using state-of-the-art extreme conditions of high magnetic fields combined with very low temperatures, in a dilution fridge in pulsed fields up to 56~T and also up to 88~T in helium-4 cryostats at the LNCMI, Toulouse \cite{pulsed1,pulsed2}, and dc fields to 33~T at the HFML, Nijmegen. Low-field measurements (up to 13.5\,T) were performed in a Quantum Design PPMS in Oxford. The Hall and resistivity contributions were separated by (anti)symmetrizing the data obtained in positive and negative magnetic fields applied along the $c$ axis. Good electrical contacts were achieved both by spot welding and Sn or In solder and to avoid heating effects electrical currents below 10~mA in pumped helium cryostats or 1~mA for the fridge were used. Quantum oscillations were observed in four samples and two samples were measured below 1.4\,K in the dilution fridge. One sample (S1) was also measured up to 88\,T at 1.4\,K.

\begin{figure}[t!]
	\centering
	\includegraphics[width=8.5cm]{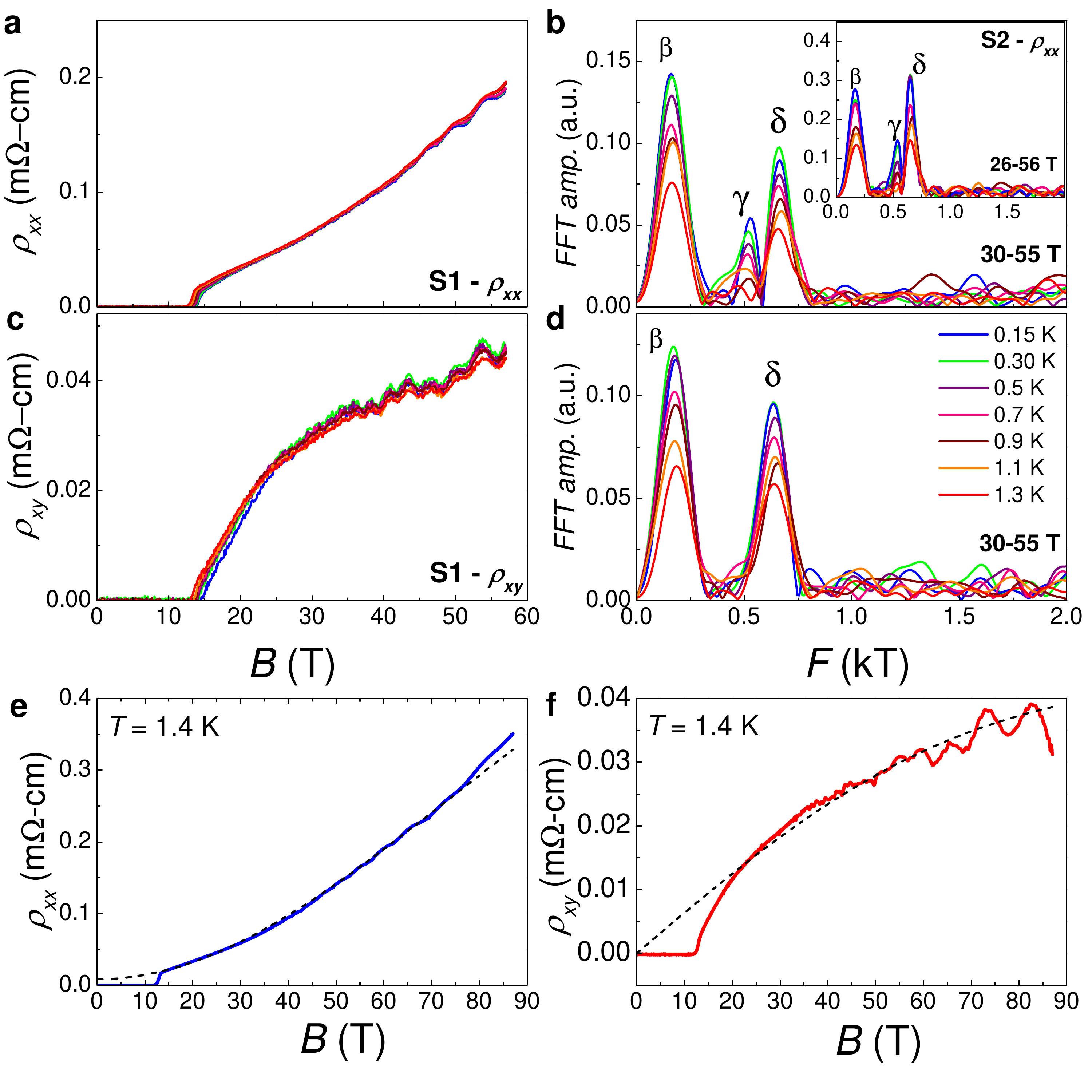}
	\caption{(color online). {\bf High-field magnetotransport data}.
		Magnetic field dependence of (a) $\rho_{xx}$ and (c) $\rho_{xy}$ for sample S1 at dilution fridge temperatures.
		The corresponding fast Fourier transform (FFT) spectra of (a) and (c) using a large field window (27-56\,T) is shown in (b) and (d), respectively, which identifies a series of frequencies, $\beta$, $\gamma$ and $\delta$. Notably, the $\gamma$ peak is absent (or very weak) in the the Hall effect component, $\rho_{xy}$. The inset shows the Fourier spectrum for sample S2. Ultra-high magnetic field (e) resistivity and (f) Hall effect measurements up to 88\,T at 1.4\,K for sample S1. The dashed lines correspond to a combined fit to a two-band model, as described in the text and Ref.\:\cite{Note1}.}
	\label{fig:FeSe_dilution}
\end{figure}

{\it High-field magnetotransport measurements.}
 Figures\:\ref{fig:FeSe_dilution}(a) and (c) show the field dependence
 of $\rho_{xx}$ and $\rho_{xy}$ for sample S1 at very low temperatures (below 1.3\,K).
We observe the transition from the superconducting
to the normal state around 14\,T,
and a very large non-saturating magnetoresistance at low temperatures. Both components of the resistivity tensor show quantum oscillations. The observation of quantum oscillations in the Hall effect
is less commonly reported than Shubnikov-de Haas oscillations in resistivity, but have been recently studied in other
multiband systems, such as cuprates \cite{Doiron2007} and Ca$_3$Ru$_2$O$_7$ \cite{Kikugawa2010}, although to the best of our knowledge not previously in Fe-based superconductors.

{\it Quantum oscillations analysis.}
The temperature-dependence  of the amplitude of quantum oscillations in resistivity are usually
well described by the Lifshitz-Kosevich formula describing
oscillations  in  thermodynamic  quantities \cite{Shoenberg}.
Analysis of the temperature-dependence of the amplitude of oscillations,
 as shown in Figs.\:\ref{fig:FeSe_dilution}(b) and (d), allows us to determine the effective masses by fitting to the Liftshitz-Kosevich formula (see Table\:S\ref{tab:freqmasses} in Ref.\:\footnote{See Supplemental Material at http://lik.aps.org/supplemental/xxx for supplemental data and discussions}).
   We find good agreement in the values of the quantum oscillations frequencies
   and effective masses between samples (within the error bars) and with those reported previously by Terashima {\it et al.} \cite{Terashima2014}. The $\gamma$ orbit has a particularly heavy effective mass of $7$-$8m_e$, whereas the $\beta$ and $\delta$ orbits have effective masses of around $4m_e$, which suggests a different origin for the two frequencies.
The field window in $1/B$  used in the pulsed field experiments (26-56\,T for sample S2) is not a large enough FFT window to identify frequencies smaller than the $\beta$ frequency at 180\,T, as compared with studies in {\it dc} fields in which an $\alpha$ pocket of about 50~T has been resolved  \cite{Terashima2014}.

Usually, it is impossible to determine based purely on quantum oscillations data whether a given quantum oscillation frequency arises from an electron-like or hole-like band, although often a comparison to band structure calculations can give a satisfactory understanding. However, here we develop a novel argument based on a compensated two carrier model and the relative strength of oscillation peaks in $\rho_{xx}$ and $\rho_{xy}$ in order to argue that the $\gamma$ peak originates from an electron-like Fermi surface pocket, whereas the $\beta$ and $\delta$ are associated with the minimum and maximum of the hole-like band.

Our approach is based on the observation that the $\gamma$ orbit is missing, or at least much weaker, in the $\rho_{xy}$ FFT spectrum than the $\rho_{xx}$, as can be directly seen by comparing the FFT spectra from sample S1 in Figs.\:\ref{fig:FeSe_dilution}(b) and (d). Note that in this sample, $\rho_{xx}$ and $\rho_{xy}$ are calculated by symmetrizing data from the same pair of contacts, which rules out any external explanation such as temperature gradients in the sample. In a multiband system, $\rho_{xx}$ and $\rho_{xy}$ may be written in terms of the conductivities of the various bands. It can be shown (see Ref.\:\cite{Note1} for derivation) that in a compensated metal (i.e. equal number of holes and electrons which stoichiometric FeSe must have) where both the electron and hole conductivities have an oscillatory component, that the Hall effect is \textit{relatively} more sensitive to the oscillations of the more mobile carrier compared to resistivity; that is to say that if both electron-like and hole-like oscillations are observed in an FFT of $\rho_{xx}$ oscillations, the FFT spectrum of $\rho_{xy}$ will show a relatively enhanced amplitude of the more mobile carrier. In FeSe, the sign of the Hall coefficient at low temperatures and high fields is positive, {\it i.e.} the holes are (on average) more mobile, therefore the hole-like orbits are enhanced in the $\rho_{xy}$ FFT spectrum and the missing / weak 540\,T $\gamma$ peak is an electron-like orbit and both the $\beta$ and $\delta$ frequencies are likely to be hole-like frequencies.

By assigning the $\delta$ and $\beta$ with the maximum and minimum of a quasi-2D pocket (as suggested in Ref.\:\cite{Terashima2014}), we estimate that this band would contain 0.014 electrons / Fe, equivalent to carrier density for the hole band of $n_h=3.6 \times 10^{20}$\,cm$^{-3}$. We note that the observation of a positive high field Hall coefficient which implies that the hole carriers are more mobile (on average) than the electron carriers is unusual in Fe-based superconductors, where the Hall effect is typically found to be negative (e.g. \cite{Kasahara2010,Kasahara2012,Rullier-Albenque2009,Albenque2012}), and quantum oscillations studies have typically found much higher amplitudes (longer scattering times) for the electron pockets than the hole pockets \cite{Coldea2008,Arnold2011a}, and sometimes the hole pockets are not observed at all \cite{Putzke2012,Shishido2010}.

\begin{figure}[t]
	\centering
\includegraphics[width=8.1cm]{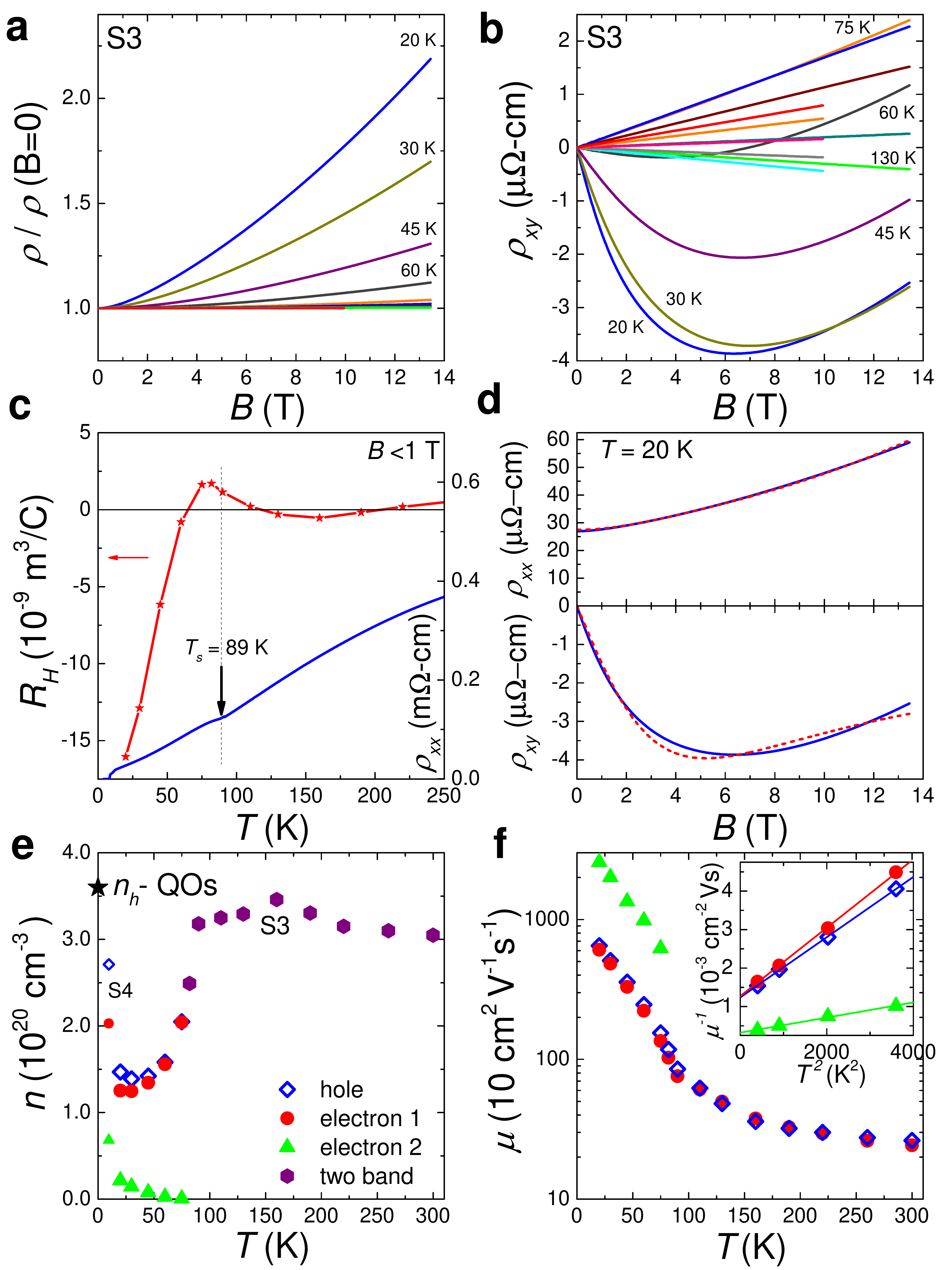}
	\caption{(color online). {\bf High-temperature magnetotransport data.} (a) Magnetoresistance measured to 13.5\,T in a sample S3 at temperatures between 20-300\,K. (b) Hall effect in FeSe. (c) low-field Hall coefficient, defined by a straight-line fit to data $<$ 1\,T.
The temperature dependence of the resistivity for S3 is also shown. (d) Simultaneous fit of the magnetoresistance and Hall effect using a constrained compensated three-carrier model. (e) Temperature-dependence of the carrier density extracted from two- and three-carrier fits (a three-carrier model is used below 75\,K which accounts
 for the development of the low-field non-linearity in the Hall effect data). (f) Temperature-dependence of mobility. Inset: the inverse of mobility, proportional to the scattering rates, which has a $\sim{}T^2$ dependence in the range of 20-60\,K.}
	\label{fig:LowField}
\end{figure}

{\it Low-field magnetotransport behavior.}
So far we have focused on the low-temperature, high magnetic field measurements, since these are the conditions where quantum oscillations are observed. However, a lot of information at higher temperatures can be obtained from low-magnetic field studies. Figures\:\ref{fig:LowField}(a) and (b) show the temperature dependence of Hall effect and the relative transverse magnetoresistance up to 14\,T of a single crystal of FeSe (S3). At high temperature, the Hall effect is observed to be linear up to 10\,T,  and
  changes sign twice above 90\,K. In this high-temperature regime, the most natural model to turn to is a compensated two-band model, where the two bands represent the electron and hole pockets
  (charge compensation is required in a stoichiometric FeSe crystal) and by fitting simultaneously $\rho_{xx}$ and $\rho_{xy}$
   we can extract three free parameters $n=n_e=n_h$, $\mu_e$ and $\mu_h$.
    A feature of this model is that the Hall effect is strictly linear at all magnetic fields and the magnetoresistance does not saturate \cite{Note1}. This model was also used for LiFeAs, where it was found that the Hall effect was negative and linear up to 14~T at all temperatures \cite{Albenque2012}.

  The temperature dependence of the extracted parameters,  assuming an equal number of holes and electrons with the carrier density $n=n_h=n_e$ and the two (field-independent) mobilities $\mu_h$ and $\mu_e$,  is shown in Figs.\:\ref{fig:LowField}(e) and (f), with the carrier density being relatively constant at high temperatures, $n \sim 3 \times 10^{20}$ cm$^{-3}$, which is also close to the low-temperature values extracted for the hole band density from quantum oscillations.  Whilst in many iron-based superconductors the electrons are generally more mobile and therefore the Hall effect is commonly found to be negative \cite{Albenque2012}, in the case of FeSe the mobilities of holes and electrons are very similar at high temperature with subtly different temperature dependencies [see Fig.\:\ref{fig:LowField}(f)]. The sign of low-field $R_H$ [Fig.\:\ref{fig:LowField}(c)] changes twice above 90\,K as at some temperatures the electrons are slightly more mobile, and at other temperatures the holes are.

{\it The three-band model.}
Below the structural transition at $T_s$ ($\sim $ 75~K), the Hall effect becomes noticeably non-linear and such behavior cannot be understood in a compensated two-band model. Recently, a mobility spectrum analysis of FeSe \cite{Huynh2014} demonstrates the presence of high mobility electron carriers (besides less mobile hole and electron carriers)  which could account for the non-linear Hall effect and, in particular, the negative sign of the Hall coefficient at low fields, but becoming positive at high fields. The Hall effect in FeSe is also similar to those on annealed BaFe$_2$As$_2$ single crystals \cite{Ishida2011}. Although the Fermi surface in the magnetically reconstructed phase is completely different there are coincidental similarities in that there is a small but compensated number of carriers and in particular a small number of highly mobile electron-like carriers. A three-carrier model \cite{Kim1999} can describe well
the Hall effect data in annealed single crystals of BaFe$_2$As$_2$ \cite{Ishida2011} and as FeSe display a similar
behavior we implement this approach but further constrained it by simultaneously fitting $\rho_{xx}$ and $\rho_{xy}$ for a compensated system with one hole
and two electron pockets (with five free parameters for mobilities and carrier concentrations). Figure\:\ref{fig:LowField}(d) at 20\,K shows the fits of this model which accounts well for the low field, non-monotonic behavior and
reveals the presence of a small but highly mobile electron pocket.
  Thus, our analysis suggests that Fermi surface of FeSe needs to have more than one electron band
  and the magnetotransport behaviour is better described if three-carriers are present, {\it i.e} one hole and two electron pockets.

Recent high-resolution ARPES data on FeSe \cite{Watson2014}, show that
the Fermi surface suffers significant deformation breaking the
rotational symmetry.
At high temperatures, the Fermi surface is composed of two hole bands ($n_{h1} \gg n_{h2}$)
and likely two electron bands ($n_{e1}$ and $n_{e2}$), which
are well compensated. At low temperatures, however, there is a single hole band $n_{h1}$
with a size close to one of the electron bands $n_{e1}$, however
the second electron band $n_{e2}$ is rather small. 
In terms of magnetotransport, this scenario is equivalent to having two types of electron-like carriers, one
with similar but slightly lower mobility to the hole bands, whereas the low-density electron-like carriers $n_{e2}$, may have lighter effective masses
and consequently higher mobility.
Using a three band model from magnetotransport data on FeSe,
 the new small pocket has a carrier density of $n_{e2}=0.7 \times 10^{20}$\,cm$^{-3}$
 and high mobility with 1843\,cm$^{2}$/Vs.
 In the limit of similar scattering, this pocket would correspond to a very low
  frequency of 90(10)T in the quantum oscillations with a lighter effective mass of $\sim 1.5(5) m_e$.
  These values are in good agreement with those extracted for this pocket from
quantum oscillations for the $\alpha$ pocket \cite{Terashima2014} and
the small electron band from high resolution ARPES studies \cite{Watson2014}.

In the low-temperature limit, in the normal state (once the superconductivity is suppressed above 14\,T) the $\rho_{xx}$ and $\rho_{xy}$ data is well-described by a two-band model over a large field window, with the extracted values listed in Table.\:\ref{tab:param}. We find that the hole mobility is 1.9-2.7 times larger than the electron mobility, which would be enough to give a noticeable difference in the relative amplitudes of oscillations in $\rho_{xx}$ and $\rho_{xy}$ as argued above. For a perfectly compensated system, $\rho_{xy}$ is exactly linear in the high field limit both in the two and three band models, whereas experimentally there is a small amount of curvature at high fields. This could indicate a small departure from stoichiometry, or possibly a field-dependence of mobility. Another possibility concerns the field-induced Fermi-surface effects caused by the Zeeman splitting whose energy scale is expected to become comparable to the effective Fermi energy of the smallest band ($\sim 2$-4\,meV \cite{Terashima2014,Kasahara2014}) in the highest magnetic field range. Although no clear signature of such a Lifshitz transition is observed, the presence
of a small pocket superimposed on other frequencies is hard to resolve.
Despite this, our quantum oscillations analysis in very high magnetic fields
based on a two-carrier model allows us to assign reliably the origin of the hole and electron pockets.

While at low fields and high temperatures the two- and three-carrier models capture the data well, the extracted temperature-dependence of the carrier density shows a marked reduction of the carrier density below  $T_s$ [Fig.\:\ref{fig:LowField}(e)], dropping by a factor of $\sim2$ compared to the high-temperature values. Furthermore, the carrier density $n_h$ estimated from quantum oscillations (assuming the $\beta$ and $\delta$ bands form the quasi-2D hole band) is $3.6\times{}10^{20}$\,cm$^{-3}$ - similar to the high-temperature carrier densities but substantially higher than the 20 K data. Below 20 K, the carrier density seems to recover, as shown by extracted parameters on a different sample S4 (see Ref.\:\cite{Note1}), and furthermore the pulsed-field low-temperature data gives carrier densities from two carrier fits which are comparable with those estimated from quantum oscillations.
In order to reconcile the apparent drop in carrier density below $T_s$ with the observation from ARPES \cite{Watson2014} and expectation that the Fermi surface should not change significantly in the absence of any spin density wave order, we speculate that a strongly anisotropic scattering rate may exist on the Fermi surfaces below the structural transition. If there are strong spin fluctuations (as observed by NMR below $T_s$ \cite{Bohmer2014}), parts of the hole and electron Fermi surfaces which are well nested by the antiferromagnetic ordering vector will have a dramatic increase in the scattering rate, similar to  calculations in Ref.\:\cite{Breitkreiz2014}, and will be effectively {\it shorted out} by non-nested segments of the Fermi surface with a much longer scattering time (and which may have conventional $\sim{}T^2$ Fermi-liquid temperature dependence of mobility [Fig.\:\ref{fig:LowField}(f)]). This may also be related to the anomalous quasiparticle interference in FeSe, which shows very strong in-plane anisotropy of energy dispersion \cite{Kasahara2014}. One can then envisage that in our simple model which assumes isotropic scattering on all bands, the drop in the \textit{effective} carrier density, despite no actual change in Fermi surface volume, could be a manifestation of strongly anisotropic scattering,
 with almost half of the available carriers {\it missing}.
However, at very low temperatures, the scattering becomes more isotropic (impurity scattering only in the $T=0$ limit) \cite{Breitkreiz2014} and in this limit (in which we observe quantum oscillations)
we find that the carrier densities extracted from the high-field two-carrier model are comparable to those estimated from quantum oscillations, $n_h$ (see Fig.\ref{fig:LowField}e).

\begin{table}[t]
\caption{Parameters extracted from magnetotransport measurements on FeSe using
 both compensated ($n_{e1}$=$n_{h1}$) and uncompensated ($n_{e1} \neq n_{h1}$)
 two-band model to fit the low-temperature-high field regime as well as
 a three-band model to fit the low field-high temperature data for different samples, as discussed in the text.
 The carrier density $n$ and mobility $\mu$ of each carrier can be compared with
 quantum oscillations value for the hole band of $n_h=3.6 \times 10^{20}$\,cm$^{-3}$. }
\begin{tabular}{llllll}
\hline
\hline
parameters (model)  & $\Delta T$ (K)  & $\Delta B$ (T)  & $h1$ & $e1$ & $e2$ \\
\hline
$n$ (2-band, S2) (10$^{20}$ cm$^{-3}$) & $1.5$ & 28-56  & 3.8(1)    & 4.7(1)       & --  \\
$\mu$ (2-band, S2) (cm$^{2}$/Vs) &  &  & 1335     & 499    &    --  \\
$n$ (2-band, S2) (10$^{20}$ cm$^{-3}$)&  $1.5$ &  28-56  & 3.5(1)    & 3.5(1)       & --  \\
$\mu$ (2-band, S2) (cm$^{2}$/Vs) & &   & 756     & 401    &    --  \\
$n$ (3-band, S4) (10$^{20}$ cm$^{-3}$) & $ 10$ & 0-13.5   & 2.71     & 2.03   &  0.68 \\
$\mu$ (3-band, S4) (cm$^{2}$/Vs)& &  & 623       & 457 & 1843 \\
\hline
\end{tabular}
\label{tab:param}
\end{table}

In summary, the magnetotransport studies in ultra-high magnetic fields and at low temperatures allow
to access the Fermi surface of FeSe and to determine independently the sign of carriers.
 In the limit of isotropic scattering, our novel analysis of the quantum oscillations in the resistivity and Hall effect,
 suggest the presence of  hole-like bands with more
  mobile carriers than the almost compensated electron-like band.
  Furthermore, the high-temperature and low-field data can be described
  only if an additional highly mobile electron pocket is considered.
  We find that the magnetotransport behavior of FeSe is a direct consequence
   of the small multiband Fermi surface with different carrier mobilities.
   Further theoretical work will be required to understand the role played by anisotropic scattering in FeSe
 and how it can affect its normal and superconducting properties.

We thank R. Arita,  A.\,E. B\"ohmer, A. Boothroyd, P.\,J. Hirschfeld, H. Ikeda, M. Rahn, T. Shimojima, M.-T. Suzuki, and T. Terashima for fruitful discussion.
This work was supported both by Japan-Germany Research Cooperative Program, KAKENHI from JSPS, and EPSRC (EP/L001772/1, EP/I004475/1, EP/I017836/1).
Part of the work was performed at the LNCMI, Toulouse, and HFML, Nijmegen, members of the
European Magnetic Field Laboratory (EMFL).
AIC acknowledges an EPSRC Career Acceleration Fellowship (EP/I004475/1).

\bibliography{FeSe_bib2}

\newpage


\newpage

\section{Supplemental Material}

\renewcommand{\tablename}{TABLE S$\!$}
\renewcommand{\figurename}{FIG. S$\!\!$}
\setcounter{table}{0}
\setcounter{figure}{0}

\begin{table}[h]
\caption{Frequencies, effective masses, areas of $k$-space and Fermi $k$-vectors for two different samples: sample S1 and sample S2  the analysis is split for the resistivity and Hall effect channels. Areas of the Brillouin Zone are relative to the high-temperature tetragonal phase.}
\begin{tabular}{llllll}
sample & orbit~  & $F$ (kT) & $m*$~($m_e$) & $A_k$ (\%BZ) & $k_F$ (\AA$^{-1}$)\\ \hline
S2: $\rho_{xx}~~~~$  &$\beta$ & 0.171    & 4.20(25)  & 0.59    & 0.072    \\
(27-56\,T)  &$\gamma$ & 0.536   & 7.55(61)   & 1.84    & 0.128    \\
~ &$\delta$ & 0.651   & 4.47(30)  & 2.24    & 0.141   \\ \hline
S1: $\rho_{xx}$  &$\beta$ & 0.164    & 4.06(23)  & 0.562    & 0.070    \\
(30-55\,T)  &$\gamma$ & 0.528   & 7.25(58)   & 1.82    & 0.127    \\
~ &$\delta$ & 0.664   & 4.38(25)  & 2.282    & 0.142   \\ \hline
S1: $\rho_{xy}$  &$\beta$ & 0.177    & 4.16(23)  & 0.609    & 0.073    \\
(30-55 T)  &$\gamma$ & -   & -   & -    & -    \\
~ &$\delta$ & 0.640   & 4.16(23)  & 3.88(25)    & 0.139   \\ \hline
\end{tabular}
\label{tab:freqmasses}
\end{table}

\subsection{The relative amplitudes of oscillations in resistivity and Hall effect in semimetals}.

\begin{figure}[htbp]
	\centering
		\includegraphics[width=8.5cm]{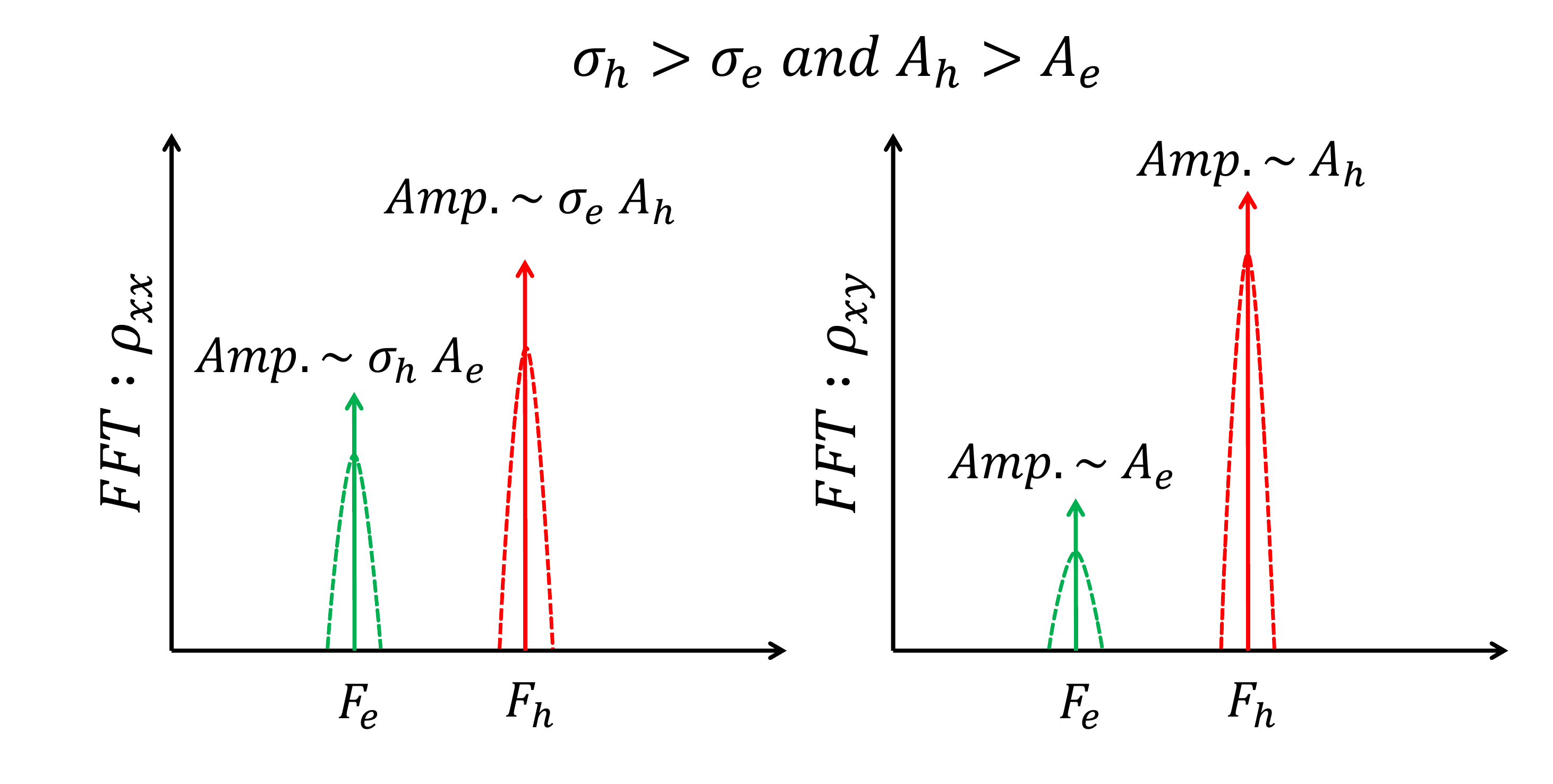}
	\caption{Cartoon of the relative amplitudes of quantum oscillations for a compensated semi-metal, with hole and electron contributions to the resistivity and Hall effect. As we have $\sigma_h > \sigma_e$ from the sign of the Hall effect, we therefore also take $A_h>A_e$. The relative amplitudes of the oscillations can therefore be understood by general considerations of a two-carrier model and suggest the nature of the carrier when the effect is large, such as in FeSe. }
	\label{fig:RelativeAmplitudesCartoon}
\end{figure}

In general, we may write the magnetoresistivity and Hall resistivity of a two-carrier system in a magnetic field $H$ as follows:

\begin{equation}
\rho_{xx}=\frac{(\sigma_h+\sigma_e)+\sigma_h\sigma_e(\sigma_hR_h^2+\sigma_eR_e^2)H^2}{(\sigma_h+\sigma_e)^2 + \sigma_h^2\sigma_e^2(R_h+R_e)^2H^2}
\end{equation}
and
\begin{equation}
\rho_{xy}=\frac{(\sigma_h^2R_h+\sigma_e^2R_e)+\sigma_h^2\sigma_e^2R_hR_e(R_h+R_e)H^2}{(\sigma_h+\sigma_e)^2 + \sigma_h^2\sigma_e^2(R_h+R_e)^2H^2}H,
\end{equation}
where $\sigma_i$ and $R_i$ are the conductivities and (inverse) carrier densities of the two species.

Since FeSe is a compensated material, we may write $R_e=-R_h=\frac{1}{ne}$ and then the formula reduce as follows:
\begin{equation}
\rho_{xx}=\frac{1+\frac{1}{(ne)^2} \sigma_h\sigma_eH^2}{(\sigma_h+\sigma_e)}
\end{equation}
and
\begin{equation}
\rho_{xy}=\frac{\frac{1}{(ne)} (\sigma_h-\sigma_e)H}{(\sigma_h+\sigma_e)}.
\end{equation}

Note that the denominators are the same in both equations. Now, we let the conductivities have an oscillatory contribution according to the Shubnikov-de Haas effect:
\begin{equation}
\sigma_i=\sigma_{i0}\left ( 1 + A_{i}\sin{ \left ( \frac{2\pi F_{i}} {B} \right )} \right ).
\end{equation}
There may be more than one oscillation frequency associated with each carrier species due to Fermi surface corrugation but we omit the summation for simplicity.

Calculating the oscillatory part of the signals, to first order in $A$ (i.e. $A<<1$), we find:

\begin{align}
\rho_{xx}^{osc} &= \frac{\sigma_{h0}\sigma_{e0}}{(ne)^2(\sigma_{h0}+\sigma_{e0})^2} \times \nonumber \\
 &\left (\sigma_{e0}A_h\sin{ \left ( \frac{2\pi F_{h}}{B} \right )} +\sigma_{h0}A_e\sin{ \left ( \frac{2\pi F_{e}} {B} \right )} \right )H^2
\end{align}
and
\begin{align}
\rho_{xy}^{osc} &= \frac{2\sigma_{h0}\sigma_{e0}}{(ne)(\sigma_{h0}+\sigma_{e0})^2} \times \nonumber \\
&\left (A_h\sin{ \left ( \frac{2\pi F_{h}}{B} \right )} -A_e\sin{ \left ( \frac{2\pi F_{e}} {B} \right )} \right )H.
\end{align}

Note in $\rho_{xx}$ the oscillation amplitude is multiplied by the mobility of the \textit{other} carrier species. Therefore for a two-carrier semimetal, the oscillations of $\rho_{xx}$ are relatively less sensitive to the oscillations of the more mobile carrier; alternatively the Hall oscillations are proportionally more sensitive to oscillations of the more mobile species, which may be determined by the sign of the Hall coefficient. E.g. in FeSe since the Hall coefficient is positive (at low temperature, in the field region of interest) we have $\sigma_h > \sigma_e$ and the oscillations of the hole bands are enhanced over the electron bands, helping to identify the oscillation frequencies. The more mobile species are likely to have a larger amplitude (greater $A_i$)  anyway as they scatter less: the scattering time entering the mobility also enters in the Dingle term determining the amplitude of the quantum oscillations, however in $\rho_{xy}$ the amplitude is further enhanced relative to the amplitude in $\rho_{xx}$. This is demonstrated schematically in Fig.\:S\ref{fig:RelativeAmplitudesCartoon}.

There are important assumptions in this formalism to be addressed here. Usually, quantum oscillations in magnetotransport are understood to occur due to oscillations of the scattering rate \cite{Shoenberg1984} according to Fermi's Golden Rule; as the density of states at the Fermi level oscillates, so too does the density of final states available after a scattering process. We have assumed that the scattering rate of the hole mobility depends on quantum oscillations arising from hole pockets, but not of electron pockets. If there is scattering that may transfer a carrier to the other carrier pocket, then
this simple model will required modification. For example, in the multiband system Ca$_3$Ru$_2$O$_7$ \cite{Kikugawa2010}
the oscillations in the Hall effect was explained by the same oscillating part of the scattering time for the two different bands, which is an opposite approach to the model presented above.
However, one may note that at low temperatures ($<4$~K) the dominant scattering process will be impurity scattering, which resembles Rutherford scattering  and strongly favors low-$q$ processes. Based on the fact that pockets in FeSe are small and well-separated in the Brillouin zone, the scattering processes will overwhelmingly leave {\it e.g.} a hole carrier in the hole band and the matrix elements for interband process are small.



\newpage
\subsection{Additional high field and high temperature magnetotransport data}

\begin{figure}[h!]
	\centering
\includegraphics[width=8cm]{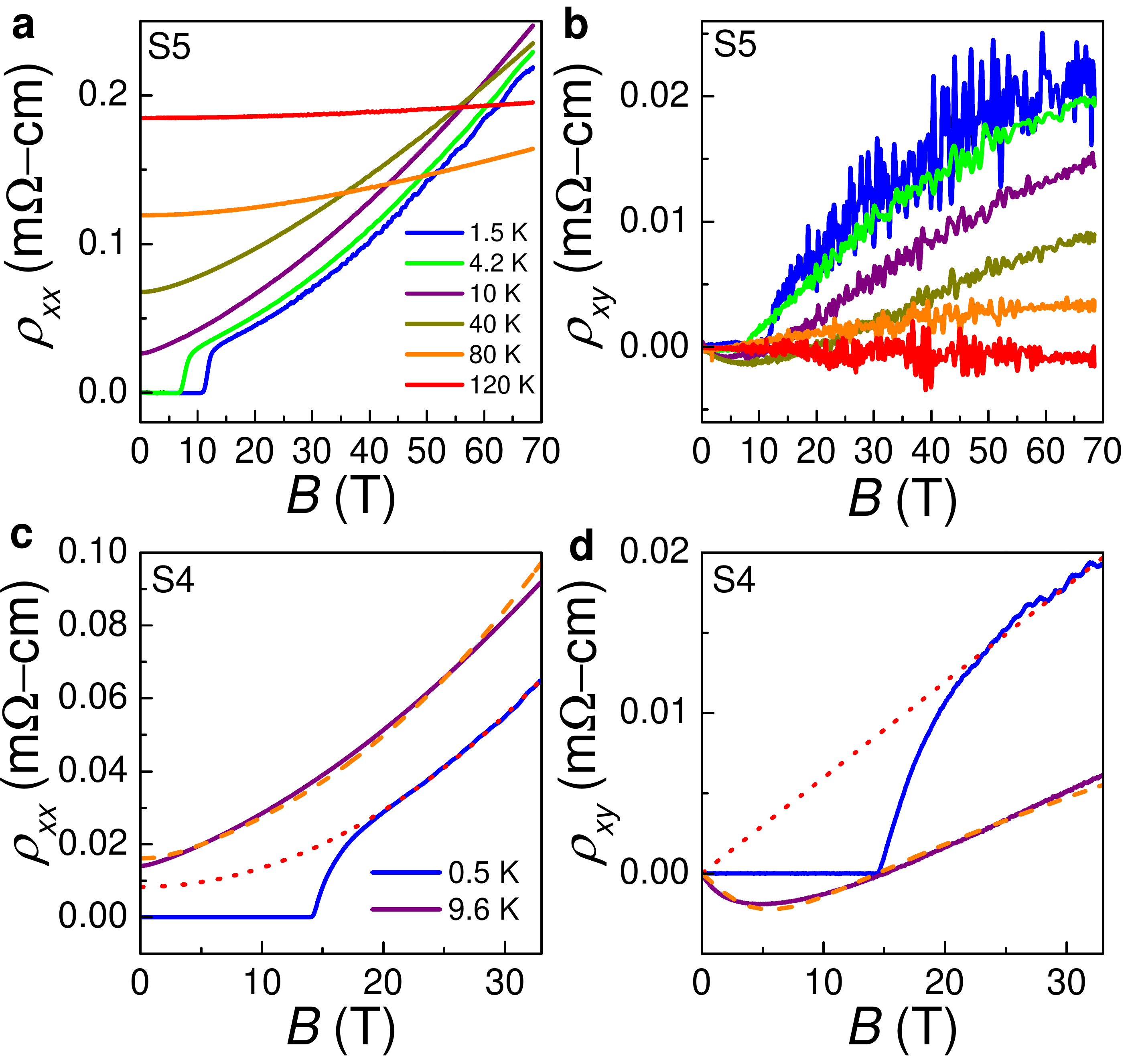}
	\caption{{High temperature magnetotransport data.} (a), (b) Pulsed field magnetotransport measurements up to 67\,T, between 1.5\,K and 120\,K. (c), (d) Measurements of the resistivity and Hall effect in dc fields for the sample S4. Orange dashed line is a fit to the 9.6\,K data using a constrained three-carrier model. Red dotted line is a fit to a constrained two-carrier model to the data above 20\,T.}
	\label{fig:highfieldSM}
\end{figure}

In Fig.\:S\ref{fig:highfieldSM}, we present further magnetoresistance data taken in pulsed and {\it dc} field magnets for different crystals. In Fig.\:S\ref{fig:highfieldSM}(a) and (b) high magnetic fields are used to measure the magnetoresistance up to 120\,K. As the mobilities drop with increasing temperature, the magnetoresistance naturally drops, until at 120\,K it is a small effect ($\Delta{}\rho (66~\rm{T})/\rho (0) = 5.5 \%$) but shows no sign of saturating at any temperature or field.
The high field Hall effect measurements on the crystal were strongly affected by noise but
 they still show that the high-field Hall effect is always positive at least up to 80\,K.

Figures.\:S\ref{fig:highfieldSM}(c) and (d) show magnetotransport measurements for another sample, S4, measured in {\it dc} fields which displays significant differences between the low-field behavior between the measurements at 0.5\,K as compared with those at 9.6\,K due to the presence of the superconductivity state. At 9.6\,K, a constrained three-carrier model describes very well the Hall effect and resistivity (orange dashed line) and the fitting parameters are listed in Table I in the main text, in agreement with the  previous low-field data in Fig.\:2 (main text). As the low field data (which is the region which allows us to detect the high mobility electron carrier) are hidden by the superconductivity
below 10\,K, a constrained two-carrier model is suitable to describe well the high field data, with the parameters
given in Table I.

%
\vspace{1cm}

\subsection{Magnets and cryogenics used in the pulsed magnetic fields experiments}

\begin{figure}[h!]
	\centering
\includegraphics[width=8cm]{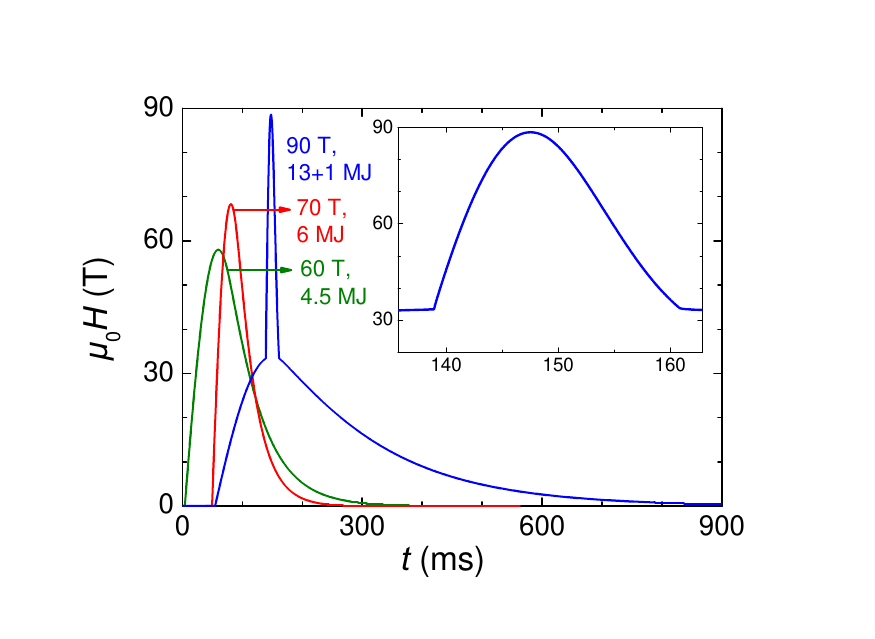}
	\caption{{\bf Ultra-high magnetic-field profile.} Temporal profile of the 60~T, 70~T and 90~T pulsed magnets, whose rising/total duration are of 55/410 ms, 31/235 ms, and (coil-exterior: 92/1100 ms; coil-interior: 8.6/22 ms), respectively. The inset shows the field up to 88~T produced by the inner magnet on top of the plateau at 33.5~T generated by the outer magnet.}
	\label{FigS3}
\end{figure}

Experiments in pulsed magnetic fields have been performed using long-duration 60~T, 70~T, and 90~T magnets at the LNCMI high-field facility in Toulouse. In particular, this work presents the first experimental study performed at helium temperature using a LNCMI non-destructive 90~T dual magnet, made of an outer coil generating 34~T over a duration of 1100~ms and an inner coil generating 56~T over a duration of 22 ms (see Fig.~\ref{FigS3}). This magnet allows experiments in excellent signal-over-noise conditions (similar to that of standard 60~T magnets) and its long duration permits to study fast variations of the physical properties. The long-duration of the 60~T magnet allowed us operating a plastic dilution fridge at temperatures down to 150~mK with no significant heating of our FeSe samples. We checked the absence of heating of the samples by Eddy currents by comparing the quantum oscillations extracted from the magnetoresistivity in the rise and fall of the pulse. Complementary details can be found in Refs.\onlinecite{pulsed1,pulsed2}, where a similar dual 80~T magnet has been used for Fermi surface studies up to 80~T  at 1.5 K \cite{pulsed1,pulsed2} and where the dilution fridge has been operated in fields up to 60~T and temperatures down to 100 mK \cite{pulsed2}.

\end{document}